\documentstyle[11pt]{article}

\begin{document}

\title{ Newton-Cartan Connections with Torsion}
\author{\Large{Tekin Dereli\footnote{E.mail: tdereli@ku.edu.tr}} \\{\it Department of Physics, Ko\c{c}
University}\\
 {\it 34450 Sar{\i}yer, \.{I}stanbul, Turkey }\\ \\
       {\Large \c{S}ahin Ko\c{c}ak\footnote{E.mail: skocak@anadolu.edu.tr}
      \, and Murat Limoncu\footnote{E.mail: mlimoncu@anadolu.edu.tr}}\\{\it Department of Mathematics,
      Anadolu University}\\
       {\it 26470 Eski\c{s}ehir, Turkey}\\
        }

\maketitle


\begin{abstract}

\noindent We  re-formulate the notion of a Newton-Cartan manifold
and clarify the compatibility conditions of a connection with
torsion with the Newton-Cartan structure.

\end{abstract}


\maketitle


\section{Introduction}

\noindent According to Einstein's theory of general relativity,
the space-time $(M, g, \nabla)$  is a four dimensional
differentiable manifold $M$ equipped with a type-(0,2), symmetric,
non-degenerate, Lorentzian signatured metric tensor $g$ and the
unique metric-compatible,  torsion-free Levi-Civita connection
$\nabla$. Elie Cartan \cite{cartan} was first to notice that
Newton's theory of gravitation can be formulated  in a
4-dimensional space-time with a corank-1  degenerate metric
\cite{friedrichs}, \cite{dombrowski}, \cite{crampin}. The physical
significance of Newton-Cartan space-times had been recognised
after the 1960's \cite{havas}, \cite{trautman}, \cite{christian}.
It is well known that  Cartan is also responsible for the concept
of space-time torsion. In what follows we clarify the notion of
Newton-Cartan manifolds and connections with torsion. In our main
theorem (Th.2) we characterize n-dimensional Newton-Cartan
space-times with torsion.

\section{Newton-Cartan connections}

\noindent {\bf Definition: 1} Let $M$ be a smooth $n$-manifold, $g$ and $%
\tau $ be (smooth) tensor fields of type $(0,2)$ and $(0,1)$, respectively, on
$M$ satisfying the following properties:

\noindent (G1) $g$ is symmetric and degenerate with $rank(g) = n-1$,

\noindent (G2) There exists a vector field $V\in {\cal X}\left( M\right) $
such that, $g(V,X)=0$ for all $X\in {\cal X}\left( M\right) $ and $\tau
(V)=1$.

\noindent We call such a triplet $(M, g, \tau)$ a Newton-Cartan manifold.

\medskip \noindent {\bf Remark:} The vector field $V$ in (G2) which is
called the time vector field is determined up to a scalar function: If $%
g_p(V_p,X_p) = 0$ and $g_p(W_p,X_p) = 0$ for all $X_p \in T_{p}M$ ($p \in M)$
and for two vectors $V_p, W_p \in T_{p}M$, then $V_p$ and $W_p$ must be
linearly dependent. Otherwise, $V_p$ and $W_p$ could be extended to a basis
of $T_{p}M$, giving $rank(g) \leq n-2$ and thus contradicting (G1).

\medskip

\noindent {\bf Convention:} Given any non-degenerate $(0,2)$-tensor $\tilde{g%
}$ on a finite dimensional vector space $V$, a vector $X\in V$
determines a
1-form $X^{\ast }$, called the $\widetilde{g}$-dual of $X$, by $X^{\ast }(Y)=%
\tilde{g}(X,Y)$ for any $Y\in V$. Likewise, a 1-form $\alpha \in
V^{\ast }$
determines a vector $\alpha ^{\ast }$, called the $\tilde{g}$-dual of $%
\alpha $, by $\tilde{g}(\alpha ^{\ast },Y)=\alpha (Y)$. We have
$(X^{\ast })^{\ast }=X$ and $(\alpha ^{\ast })^{\ast }=\alpha $.
We will be using also
the dual base $(f^{1},\dots ,f^{n})$ of $V^{\ast }$ corresponding to a base $%
(e_{1},\dots ,e_{n})$ of $V$, given by $f^{i}(e_{j})=\delta
_{j}^{i}$, which is to be distinguished from the $\tilde{g}$-dual
base $(e_{1}^{\ast },\dots ,e_{n}^{\ast })$.

\medskip

\noindent We will consider a fixed vector field $V$ with $g(V,X)=0$ and $%
\tau (V)=1$ also as part of a Newton-Cartan manifold. $g$ restricted to the
kernel of $\tau $ ( more precisely to $ker\tau \times ker\tau $) is
non-degenerate and one easily obtains:

\noindent {\bf Proposition:1} The tensor field $\bar{g}_{\epsilon
}=g+\epsilon \tau \otimes \tau $, where $\epsilon $ is a non-zero real
number, is symmetric and non-degenerate. Thus it is a semi-Riemannian metric
on $M$. The dual $\tau ^{\ast }$ of $\tau $ with respect to $\bar{g}%
_{\epsilon }$ is $\frac{1}{\epsilon }V$, i.e. $\tau (X)=\bar{g}_{\epsilon }(%
\frac{1}{\epsilon }V,X)$.

\noindent {\bf Remark:} In the following, unless necesary, we will drop the
subscript $\epsilon $ and understand that a fixed $\epsilon $ is chosen. We
will stress the role of $\epsilon $ in Theorems 1 and 2.

\medskip

\noindent Let us define $\bar{h}(\alpha,\beta) = \bar{g}(\alpha^{*},%
\beta^{*})$, where $\alpha$ and $\beta$ are 1-forms and $\alpha^{*} ,
\beta^{*}$ their $\bar{g}$-duals. The $(2,0)$-tensor field $\bar{h}$ is
symmetric and non-degenerate.

\medskip

\noindent {\bf Proposition: 2} The $(2,0)$-tensor field $h=\bar{h}-\frac{1}{%
\epsilon }V\otimes V$ is symmetric and degenerate.

\noindent Proof: $h$ is obviously symmetric. For degeneracy, let $\theta$ be
any 1-form field. Then
\begin{eqnarray}
h(\tau,\theta) &=& \bar{h}(\tau,\theta) - \frac{1}{\epsilon} V(\tau)
V(\theta) \quad \quad \quad \quad  \nonumber \\
&=& \bar{g}(\tau^{*},\theta^{*}) - \frac{1}{\epsilon} \tau(V) \theta(V)
\nonumber \\
&=& \bar{g}(\frac{1}{\epsilon}V,\theta^{*}) - \frac{1}{\epsilon} \theta(V)
\nonumber \\
&=& \frac{1}{\epsilon} ( \bar{g}(V,\theta^{*}) - \theta(V) ) = 0 .  \nonumber
\end{eqnarray}

\medskip

\noindent {\bf Proposition:3} The following relations between $\bar{h}$ and $%
\bar{g}$ (resp. $h$ and $g$) hold:
\[
c_1^2 (\bar{h} \otimes \bar{g}) = \delta \, ,
\]
\[
c_1^2 (h \otimes g) = \delta - V \otimes \tau
\]

\noindent where ${c_1}^2$ denotes the $(2,1)$-contraction of the $(2,2)$%
-tensor $\bar{h} \otimes \bar{g}$ (resp. $h \otimes g$) and $\delta$ is the $%
(1,1)$-tensor $\delta(\alpha,X) = \alpha(X)$. In a local coordinate chart $%
(x^1, \dots,x^n)$ we have in terms of components
\[
\sum_{k} {\bar{h}}^{ik} {\bar{g}}_{kj} = \delta^{i}_{j} ,
\]
\[
\sum_{k} {h}^{ik} {g}_{kj} = \delta^{i}_{j} - V^i \tau_j .
\]
These equalities are also true for any non-coordinate base $\{e_i\}$ and its
dual base $\{f^j\}$ with $f^j(e_i) = \delta^j_i.$

\medskip

\noindent Proof: Let $(\partial_1,\dots,\partial_n)$ denote the coordinate
base, $(dx^1,\dots,dx^n)$ the dual base. Since $\bar{g}(\partial_i,%
\partial_j) = {\bar{g}}_{ij}$ and $\bar{h}(dx^i,dx^j) = {\bar{h}}^{ij}$ we
have by definition $\sum_{k} {\bar{h}}^{ik} {\bar{g}}_{kj} = \delta^{i}_{j}.$
Inserting in this $\bar{h}^{ik} = h^{ik} + \frac{1}{\epsilon} \tau^i \tau^k$
and $\bar{g}_{kj} = g_{kj} + \epsilon V_k V_j$ and taking into account $%
\tau(V) = 1$, $h(dx^i,\tau) = 0$ , $g(V,\partial_j) = 0$ we get $\sum_k
h^{ik}g_{kj} = \delta^{i}_{j} - \tau^i V_j$.

\medskip

\noindent Having defined a semi-Riemannian manifold $(M,\bar{g})$ associated
with the Newton-Cartan manifold $(M,g,\tau )$, we have the standard theory
of connections at our disposal. Imposing certain conditions on connections
on $M$, we will single out {\sl Newton-Cartan} connections on $(M,g,\tau )$
compatible with $g$ and $\tau $ in the following sense:

\noindent {\bf Definition: 2} Let $\nabla _{X}$ denote the covariant
derivative of any linear connection ${\cal D}$ on the manifold $M$. We will
call the connection ${\cal D}$ a {\sl Newton-Cartan }connection iff
\[
\nabla _{X}\tau =0   \quad , \quad  \nabla _{X}g=0  \quad \forall X \in {\cal X}%
\left( M\right)
\]

\noindent {\bf Proposition: 4} Let $\nabla _{X}$ denote the
covariant derivative of a {\sl Newton-Cartan }connection on $M$.
Then $\nabla _{X}V=0$ and $\nabla _{X}h=0$, where $V$ is the time
vector field and $h$ the
(2,0)-field associated with $g$ (s. Prop 2). Moreover, $\nabla _{X}\overline{%
h}=0$.

\noindent Proof: First we note that $\nabla _{X}\bar{g}=\nabla
_{X}g+\epsilon \left( \nabla _{X}\tau \otimes \tau +\tau \otimes \nabla
_{X}\tau \right) =0$. Now, from Prop.1, $\epsilon \tau (Y)=\bar{g}(V,Y)$ for
any $Y\in {\cal X}\left( M\right) $. Then
\begin{eqnarray*}
\nabla _{X}(\epsilon \tau (Y)) &=&\nabla _{X}(\bar{g}(V,Y)) \\
\epsilon (\nabla _{X}\tau )(Y)+\epsilon \tau (\nabla _{X}Y) &=&(\nabla _{X}%
\bar{g})(V,Y)+\bar{g}(\nabla _{X}V,Y)+\bar{g}(V,\nabla _{X}Y) \\
\epsilon \tau (\nabla _{X}Y) &=&\bar{g}(\nabla _{X}V,Y)+\bar{g}(V,\nabla
_{X}Y)
\end{eqnarray*}
by $\nabla _{X}\tau =0$ and $\nabla _{X}\bar{g}=0$. Since $\epsilon \tau
(\nabla _{X}Y)=\bar{g}\left( V,\nabla _{X}Y\right) $ we get $\bar{g}(\nabla
_{X}V,Y)=0$, hence $\nabla _{X}V=0$.

\noindent Now we consider $h=\bar{h}-\frac{1}{\epsilon }V\otimes V$%
\begin{eqnarray*}
\nabla _{X}h &=&\nabla _{X}\bar{h}-\frac{1}{\epsilon }V\otimes \nabla _{X}V-%
\frac{1}{\epsilon }\nabla _{X}V\otimes V \\
\nabla _{X}h &=&\nabla _{X}\bar{h}
\end{eqnarray*}
So it is enough to show $\nabla _{X}\bar{h}=0$. Let $\alpha ,\beta $ be
1-form fields and $\alpha ^{\ast },\beta ^{\ast }$ their $\bar{g}$-dual
vector fields. Since by definition $\bar{h}(\alpha ,\beta )=\bar{g}(\alpha
^{\ast },\beta ^{\ast })$,
\[
(\nabla _{X}\bar{h})(\alpha ,\beta )=\nabla _{X}(\bar{h}(\alpha ,\beta ))-%
\bar{h}(\nabla _{X}\alpha ,\beta )-\bar{h}(\alpha ,\nabla _{X}\beta ).
\]
On the other hand we have $(\nabla _{X}\alpha )^{\ast }=\nabla _{X}(\alpha
^{\ast })$:
\begin{eqnarray}
\bar{g}(\alpha ^{\ast },U) &=&\alpha (U)\quad \quad U\in {\cal X}%
\left( M\right)  \nonumber \\
\nabla _{X}(\bar{g}(\alpha ^{\ast },U)) &=&\nabla _{X}(\alpha (U))  \nonumber
\\
\bar{g}(\nabla _{X}\alpha ^{\ast },U)+\bar{g}(\alpha ^{\ast },\nabla _{X}U)
&=&(\nabla _{X}\alpha )(U)+\alpha (\nabla _{X}U)  \nonumber \\
\bar{g}(\nabla _{X}\alpha ^{\ast },U) &=&(\nabla _{X}\alpha )(U)  \nonumber
\end{eqnarray}
because $\bar{g}(\alpha ^{\ast },\nabla _{X}U)=\alpha (\nabla _{X}U)$. This
means $(\nabla _{X}\alpha )^{\ast }=\nabla _{X}\alpha ^{\ast }$. Inserting
these correspondences above and from $\nabla _{X}\bar{g}=0$ we obtain
\[
\left( \nabla _{X}\bar{h}\right) (\alpha ,\beta )=\nabla _{X}(\bar{g}(\alpha
^{\ast },\beta ^{\ast }))-\bar{g}(\nabla _{X}\alpha ^{\ast },\beta ^{\ast })-%
\overline{g}(\alpha ^{\ast },\nabla _{X}\beta ^{\ast })=0  .
\]

\noindent {\bf Remark: }The converse of this proposition is also true and a
kind of duality emerges. One could define a {\sl Newton-Cartan} manifold
also as a triplet ($M,h,V$).

The following theorem clarifies the existence and uniqueness of
torsion-free {\sl Newton-Cartan} connections:

\noindent {\bf Theorem:1} Let ($M,g,\tau $) be a {\sl Newton-Cartan}
manifold.

\noindent i)\ If ${\cal D}$\ is a {\sl Newton-Cartan }connection\ without\
torsion, then ${\cal D}$\ is the Levi-Civita connection on ($M,\overline{g}%
_{\varepsilon }$) for all $\epsilon $ ($\neq 0$) and $L_{V}g=0,d\tau =0$
where $L_{V}$ denotes the Lie derivative, $V$ being the time vector field $%
(\tau (V)=1)$.

\noindent ii) Let ${\cal D}$ denote the Levi-Civita connection on ($M,%
\overline{g}_{\varepsilon }$) for a fixed $\epsilon $ ($\neq 0$). Assume, $%
L_{V}g=0$ and $d\tau =0$. Then, ${\cal D}$\ is a torsion-free {\sl %
Newton-Cartan }connection on ($M,g,\tau $).

\noindent {\bf Remark:} Theorem 1 shows that, there can't be any torsion-free
{\sl Newton-Cartan }connection, unless $L_{V}g=0$ and $d\tau =0$. In case
of $L_{V}g=0$ and $d\tau =0$, there is a unique {\sl Newton-Cartan }%
connection without torsion; it is the Levi-Civita connection for any $\bar{g}%
_{\epsilon }$ (especially, Levi-Civita connections for $\bar{g}_{\epsilon }$
coincide for all $\epsilon \neq 0$).

\medskip

\noindent Proof (of theorem 1):

i) The first assertion (that $\nabla _{X}\overline{g}_{\varepsilon }=0$) is
trivial. So we show $d\tau =0$ and $L_{V}g=0$.
\begin{eqnarray*}
\left( \nabla _{X}\tau \right) \left( Y\right) &=&\nabla _{X}\left( \tau
\left( Y\right) \right) -\tau \left( \nabla _{X}Y\right) =0 \\
\left( \nabla _{Y}\tau \right) \left( X\right) &=&\nabla _{Y}\left( \tau
\left( X\right) \right) -\tau \left( \nabla _{Y}X\right) =0
\end{eqnarray*}
\begin{eqnarray*}
X\tau \left( Y\right) -Y\tau \left( X\right) &=&\tau \left( \nabla
_{X}Y-\nabla _{Y}X\right) =\tau \left( \left[ X,Y\right] \right) \\
\left( d\tau \right) \left( X,Y\right) &=&0 \quad \quad d\tau =0 .
\end{eqnarray*}
Now we will show $L_{V}g=0$:
\begin{eqnarray*}
\left( \nabla _{X}g\right) \left( Y,Z\right) &=&\nabla _{X}\left( g\left(
Y,Z\right) \right) -g\left( \nabla _{X}Y,Z\right) -g\left( Y,\nabla
_{X}Z\right) =0 \\
&=&Xg\left( Y,Z\right) -g\left( \left[ X,Y\right] +\nabla _{Y}X,Z\right)
-g\left( Y,\left[ X,Z\right] +\nabla _{Z}X\right) =0 \\
&=&Xg\left( Y,Z\right) -g\left( L_{X}Y,Z\right) -g\left( Y,L_{X}Z\right)
-g\left( \nabla _{Y}X,Z\right) -g\left( Y,\nabla _{Z}X\right) =0
\end{eqnarray*}
Take $X=V$:
\[
Vg\left( Y,Z\right) -g\left( L_{V}Y,Z\right) -g\left( Y,L_{V}Z\right)
-g\left( \nabla _{Y}V,Z\right) -g\left( Y,\nabla _{Z}V\right) =0
\]
By Prop.4, $\nabla _{Y}V=0$ $\nabla _{Z}V=0$ and thus
\[
Vg\left( Y,Z\right) -g\left( L_{V}Y,Z\right) -g\left( Y,L_{V}Z\right) =0%
\Longrightarrow \left( L_{V}g\right) \left( Y,Z\right) =0%
 (  L_{V}g=0 ) .
\]

ii) It is enough to show $\nabla _{X}\tau =0$ as $\nabla _{X}g=0$ is then a
consequence of $\overline{g}_{\epsilon }=g+\epsilon \tau \otimes \tau $ and $%
\nabla _{X}\overline{g}_{\epsilon }=0$. (We will write below $\overline{g}$
for $\overline{g}_{\epsilon }$ again)

Let $Y\in {\cal X}\left( M\right) $. We get
\begin{eqnarray*}
\left( \nabla _{X}\tau \right) \left( Y\right) &=&\nabla _{X}\left( \tau
\left( Y\right) \right) -\tau \left( \nabla _{X}Y\right) \\
\left( \nabla _{X}\tau \right) \left( Y\right) &=&X\left( \overline{g}\left(
\frac{1}{\epsilon }V,Y\right) \right) -\overline{g}\left( \frac{1}{%
\epsilon }V,\nabla _{X}Y\right) \\
\epsilon \left( \nabla _{X}\tau \right) \left( Y\right) &=&X\left( \overline{%
g}\left( V,Y\right) \right) -\overline{g}\left( V,\nabla _{X}Y\right)
\end{eqnarray*}
since $\frac{1}{\epsilon }V$ is $\overline{g}$-dual of $\tau $. From the
Koszul formula
\begin{eqnarray*}
2\overline{g}\left( \nabla _{X}Y,V\right) &=&X\overline{g}\left( Y,V\right)
+Y\overline{g}\left( V,X\right) -V\overline{g}\left( X,Y\right) \\
&&-\overline{g}\left( X,\left[ Y,V\right] \right) +\overline{g}\left( Y,%
\left[ V,X\right] \right) +\overline{g}\left( V,\left[ X,Y\right] \right)
\qquad \qquad
\end{eqnarray*}
we can write
\begin{equation}
\begin{array}{l}
\epsilon \left( \nabla _{X}\tau \right) \left( Y\right) =\frac{1}{2}\{X%
\overline{g}\left( V,Y\right) -Y\overline{g}\left( V,X\right) -\overline{g}%
\left( V,\left[ X,Y\right] \right) \} \\
\qquad \qquad \quad \quad +\frac{1}{2}\{V\overline{g}\left( X,Y\right) -%
\overline{g}\left( X,\left[ V,Y\right] \right) -\overline{g}\left( Y,\left[
V,X\right] \right) \} .
\end{array}
\end{equation}
The first term vanishes because of $d\tau =0$:
\begin{eqnarray*}
\left( d\tau \right) \left( X,Y\right) &=&X\left( \tau \left( Y\right)
\right) -Y\left( \tau \left( X\right) \right) -\tau \left( \left[ X,Y\right]
\right) =0 \\
&=&X\left( \overline{g}\left( \frac{1}{\epsilon }V,Y\right) \right)
-Y\left( \overline{g}\left( \frac{1}{\epsilon }V,X\right) \right) -%
\overline{g}\left( \frac{1}{\epsilon }V,\left[ X,Y\right] \right) =0 .
\end{eqnarray*}
The second term in (1) vanishes because of $L_{V}\overline{g}=0$:
(we will instantly show that $L_{V}\overline{g}=0$)
\begin{eqnarray*}
\left( L_{V}\overline{g}\right) \left( X,Y\right) &=&V\overline{g}\left(
X,Y\right) -\overline{g}\left( L_{V}X,Y\right) -\overline{g}\left(
X,L_{V}Y\right) =0 \\
&=&V\overline{g}\left( X,Y\right) -\overline{g}\left( X,\left[ V,Y\right]
\right) -\overline{g}\left( Y,\left[ V,X\right] \right) =0
\end{eqnarray*}
Now the missing last part of the proof: $L_{V}\overline{g}=0$. First we show
that $L_{V}\tau =0$:
\begin{eqnarray*}
\left( L_{V}\tau \right) \left( X\right) &=&L_{V}\left( \tau \left( X\right)
\right) -\tau \left( L_{V}X\right) \\
&=&V\left( \tau \left( X\right) \right) -\tau \left( \left[ V,X\right]
\right) \\
&=&X\left( \tau \left( V\right) \right)  \qquad  ( d\tau =0 )
\\
&=&0 \qquad \qquad \qquad ( \tau \left( V\right) =1 )
\end{eqnarray*}
Now,
\begin{eqnarray*}
L_{V}\overline{g} &=&L_{V}g+\epsilon L_{V}\left( \tau \otimes \tau \right) \\
L_{V}\overline{g} &=&L_{V}g+\epsilon L_{V}\tau \otimes \tau +\epsilon \tau
\otimes L_{V}\tau \\
L_{V}\overline{g} &=&L_{V}g=0
\end{eqnarray*}
by assumption of the theorem.

\bigskip

\noindent We generalize to the cases with torsion and prove our
main theorem below:

\noindent {\bf Theorem:2} Let ($M,g,\tau $) be a {\sl Newton-Cartan} manifold

\noindent i) If ${\cal D}$\ is a {\sl Newton-Cartan }connection\ with\
torsion $T$ ($T(X,Y)=\nabla _{X}Y-\nabla _{Y}X-[X,Y]$)\ then\ ${\cal D}$\ is
$\overline{g}_{\epsilon }$ compatible for all $\epsilon $ ($\neq 0$) (i.e. $%
\nabla _{X}\overline{g}_{\epsilon }=0$) and
\begin{equation}
\begin{array}{l}
(L_{V}g)(X,Y)=g(T(V,X),Y)+g(X,T(V,Y)) \\
(d\tau )(X,Y)=\tau (T(X,Y))
\end{array}
\end{equation}
for all $X,Y,Z\in {\cal X}\left( M\right) .$

\noindent ii) Let ${\cal D}$ denote the connection on ($M,\overline{g}%
_{\epsilon }$), compatible with $\overline{g}_{\epsilon }$ and with torsion $%
T$ (for a fixed $\epsilon \neq 0$). Assume
\[
\begin{array}{l}
(L_{V}g)(X,Y)=g(T(V,X),Y)+g(X,T(V,Y)) \\
(d\tau )(X,Y)=\tau (T(X,Y))
\end{array}
\]
for all $X,Y,Z\in {\cal X}\left( M\right) $. Then ${\cal D}$\ is a {\sl %
Newton-Cartan }connection on ($M,g,\tau $) with torsion $T.$

\noindent {\bf Remark:} We call the conditions (2) {\sl compatibility conditions}
of a torsion with the {\sl Newton-Cartan} structure. Theorem 2 shows that, a
{\sl Newton-Cartan }connection with torsion $T$ can exist, only if the
torsion $T$ is compatible with the {\sl Newton-Cartan} structure. On the
other hand, if $T$ is compatible with the {\sl Newton-Cartan} structure,
then there exists a unique {\sl Newton-Cartan }connection with torsion $T$;
it is the connection compatible with any $\overline{g}_{\varepsilon }$ and
having torsion $T$ (especially, connections compatible with $\overline{g}%
_{\varepsilon }$ and having torsion $T$ coincide for all $\epsilon \neq 0$).

\medskip

\noindent Proof (of theorem 2)

i) The first assertion (that $\nabla _{X}\overline{g}_{\epsilon }=0$) is
again trivial. So we show the other two identities. First we show $(d\tau
)(X,Y)=\tau (T(X,Y))$:
\begin{eqnarray*}
\left( \nabla _{X}\tau \right) \left( Y\right) &=&\nabla _{X}\left( \tau
\left( Y\right) \right) -\tau \left( \nabla _{X}Y\right) =0 \\
\left( \nabla _{Y}\tau \right) \left( X\right) &=&\nabla _{Y}\left( \tau
\left( X\right) \right) -\tau \left( \nabla _{Y}X\right) =0
\end{eqnarray*}
\begin{eqnarray*}
X\tau \left( Y\right) -Y\tau \left( X\right) &=&\tau \left( \nabla
_{X}Y-\nabla _{Y}X\right) =\tau \left( T\left( X,Y\right) +\left[ X,Y\right]
\right) \\
\left( d\tau \right) \left( X,Y\right) &=&X\tau \left( Y\right) -Y\tau
\left( X\right) -\tau \left( \left[ X,Y\right] \right) =\tau \left( T\left(
X,Y\right) \right) .
\end{eqnarray*}
Now we show the identity about $L_{V}g$:
\[
\left( \nabla _{X}g\right) \left( Y,Z\right) =\nabla _{X}\left( g\left(
Y,Z\right) \right) -g\left( \nabla _{X}Y,Z\right) -g\left( Y,\nabla
_{X}Z\right) =0
\]
\[
Xg\left( Y,Z\right) -g\left( T\left( X,Y\right) +\left[ X,Y\right] +\nabla
_{Y}X,Z\right) -g\left( Y,T\left( X,Z\right) +\left[ X,Z\right] +\nabla
_{Z}X\right) =0
\]
\[
\begin{array}{l}
Xg\left( Y,Z\right) -g\left( L_{X}Y,Z\right) -g\left( Y,L_{X}Z\right) \\
-g\left( T\left( X,Y\right) +\nabla _{Y}X,Z\right) -g\left( Y,T\left(
X,Z\right) +\nabla _{Z}X\right) =0
\end{array}
\]
Take $X=V$:
\[
Vg\left( Y,Z\right) -g\left( L_{V}Y,Z\right) -g\left( Y,L_{V}Z\right)
-g\left( T\left( V,Y\right) +\nabla _{Y}V,Z\right) -g\left( Y,T\left(
V,Z\right) +\nabla _{Z}V\right) =0
\]
By Prop.4, $\nabla _{Y}V=0$ $\nabla _{Z}V=0$ and thus
\begin{eqnarray*}
Vg\left( Y,Z\right) -g\left( L_{V}Y,Z\right) -g\left( Y,L_{V}Z\right)
&=&g\left( T\left( V,Y\right) ,Z\right) +g\left( Y,T\left( V,Z\right) \right)
\\
(L_{V}g)(Y,Z) &=&g\left( T\left( V,Y\right) ,Z\right) +g\left( Y,T\left(
V,Z\right) \right) .
\end{eqnarray*}

ii) Again, it is enough to show $\nabla _{X}\tau =0$ as $\nabla _{X}g=0$
follows from $\overline{g}=g+\epsilon \tau \otimes \tau $ and $\nabla _{X}%
\overline{g}=0.$

The equality for $\nabla _{X}\tau $,

\[
\epsilon \left( \nabla _{X}\tau \right) \left( Y\right) =X\left( \overline{g}%
\left( V,Y\right) \right) -\overline{g}\left( V,\nabla _{X}Y\right)
\]
is still true. But now we have the Koszul formula with torsion:
\begin{eqnarray*}
2\overline{g}\left( \nabla _{X}Y,V\right) &=&X\overline{g}\left( Y,V\right)
+Y\overline{g}\left( V,X\right) -V\overline{g}\left( X,Y\right) \\
&&-\overline{g}\left( X,\left[ Y,V\right] \right) +\overline{g}\left( Y,%
\left[ V,X\right] \right) +\overline{g}\left( V,\left[ X,Y\right] \right)
\qquad \qquad \\
&&-\overline{g}\left( X,T\left( Y,V\right) \right) +\overline{g}\left(
Y,T\left( V,X\right) \right) +\overline{g}\left( V,T\left( X,Y\right)
\right) \qquad \qquad
\end{eqnarray*}
where $T\left( X,Y\right) =\nabla _{X}Y-\nabla _{Y}X-[X,Y].$

Inserting this into the equality for $\nabla _{X}\tau $ we get
\begin{equation}
\begin{array}{l}
2\epsilon \left( \nabla _{X}\tau \right) \left( Y\right) =\{X\overline{g}%
\left( V,Y\right) -Y\overline{g}\left( V,X\right) -\overline{g}\left( V,%
\left[ X,Y\right] \right) \} \\
\qquad \qquad \quad \quad +\{V\overline{g}\left( X,Y\right) -\overline{g}%
\left( X,\left[ V,Y\right] \right) -\overline{g}\left( Y,\left[ V,X\right]
\right) \} \\
\qquad \qquad \quad \quad +\{\overline{g}\left( X,T\left( Y,V\right) \right)
-\overline{g}\left( Y,T\left( V,X\right) \right) -\overline{g}\left(
V,T\left( X,Y\right) \right) \}.
\end{array}
\end{equation}
We first simplify the first term in (3):

By assumption $\left( d\tau \right) \left( X,Y\right) =\tau \left( T\left(
X,Y\right) \right) $%
\begin{eqnarray*}
X\left( \tau \left( Y\right) \right) -Y\left( \tau \left( X\right) \right)
-\tau \left( \left[ X,Y\right] \right) &=&\tau \left( T\left( X,Y\right)
\right) \\
X\left( \overline{g}\left( \frac{1}{\epsilon }V,Y\right) \right) -Y\left(
\overline{g}\left( \frac{1}{\epsilon }V,X\right) \right) -\overline{g}%
\left( \frac{1}{\epsilon }V,\left[ X,Y\right] \right) &=&\overline{g}\left(
\frac{1}{\epsilon }V,T\left( X,Y\right) \right)
\end{eqnarray*}
Thus, the first term in (3) equals $\overline{g}\left( V,T\left(
X,Y\right) \right) .$

To simplify the second term in (3) we use as before,
\begin{eqnarray*}
\left( L_{V}\overline{g}\right) \left( X,Y\right) &=&V\overline{g}\left(
X,Y\right) -\overline{g}\left( L_{V}X,Y\right) -\overline{g}\left(
X,L_{V}Y\right) \\
&=&V\overline{g}\left( X,Y\right) -\overline{g}\left( X,\left[ V,Y\right]
\right) -\overline{g}\left( Y,\left[ V,X\right] \right)
\end{eqnarray*}
Thus, the second term in (3) equals $\left( L_{V}\overline{g}%
\right) \left( X,Y\right) $. Let us compute $L_{V}\overline{g}$: Since $%
\overline{g}=g+\epsilon \tau \otimes \tau $%
\[
L_{V}\overline{g}=L_{V}g+\epsilon L_{V}\tau \otimes \tau +\epsilon \tau
\otimes L_{V}\tau
\]
so we need $L_{V}\tau $:
\begin{eqnarray*}
\left( L_{V}\tau \right) \left( X\right) &=&L_{V}\left( \tau \left( X\right)
\right) -\tau \left( L_{V}X\right) \\
&=&V\left( \tau \left( X\right) \right) -\tau \left( \left[ V,X\right]
\right) \\
&=&X\left( \tau \left( V\right) \right) +\tau \left( T\left( V,X\right)
\right) \qquad \left( d\tau \right) \left( V,X\right) =\tau
\left( T\left( V,X\right) \right) ) \\
&=&\tau \left( T\left( V,X\right) \right)  \qquad \qquad \qquad \qquad
 ( \tau \left( V\right) =1 )
\end{eqnarray*}
Inserting this in the formula for $L_{V}\overline{g}$, we obtain
\begin{eqnarray*}
\left( L_{V}\overline{g}\right) \left( X,Y\right) &=&\left( L_{V}g\right)
\left( X,Y\right) +\epsilon \left( L_{V}\tau \right) \left( X\right) \tau
\left( Y\right) +\epsilon \tau \left( X\right) \left( L_{V}\tau \right)
\left( Y\right) \\
&=&g(T(V,X),Y)+g(X,T(V,Y)) \\
&&+\epsilon \tau \left( T\left( V,X\right) \right) \tau \left( Y\right)
+\epsilon \tau \left( X\right) \tau \left( T\left( V,Y\right) \right) \\
&=&\overline{g}(T(V,X),Y)+\overline{g}(X,T(V,Y))
\end{eqnarray*}
If we now replace the first term in (3) by $\overline{g}%
(V,T(X,Y)) $ and second term in (3) by $\overline{g}(T(V,X),Y)+%
\overline{g}(X,T(V,Y))$ we get $\nabla _{X}\tau =0$ (by anti-symmetry of $T$%
).

\noindent {\bf Remark:} There are two cases where the conditions of Theorem
2 simplify. Namely,

i. If $T(X,Y) \in ker\tau$, then the condition $(d\tau)(X,Y) = \tau(T(X,Y))$
reduces to $d\tau = 0$.

ii. If $T(X,Y)=d\tau (X,Y)V$ then the condition $(d\tau )(X,Y)=\tau (T(X,Y))$
is satisfied. Because, $\tau (T(X,Y))=\tau (d\tau (X,Y)V)=d\tau (X,Y)\tau
(V)=d\tau (X,Y)$ by $\tau (V)=1.$ In this case, the condition $%
(L_{V}g)(X,Y)=g(T(V,X),Y)+g(X,T(V,Y))$ reduces to $L_{V}g=0$ because $%
g(T(V,X),Y)=g(d\tau (V,X)V,Y)=d\tau (V,X)g(V,Y)=0$ and similarly $%
g(X,T(V,Y))=0$.

Emerging from the considerations in Theorem 1 and Theorem 2 (and remarks
following them), there holds the following characterization of {\sl %
Newton-Cartan }connections, which is interesting by itself.

\medskip \noindent {\bf Proposition 5:} A connection ${\cal D}$ (with or
without torsion) on a {\sl Newton-Cartan }manifold ($M,g,\tau $) is a {\sl %
Newton-Cartan }connection, if and only if it is compatible with all metrics $%
\overline{g}_{\epsilon }=g+\epsilon \tau \otimes \tau $ ($\epsilon \neq 0$).

\noindent (i.e. $\nabla _{X}g=0$ and $\nabla _{X}\tau =0$ for all $X\in
{\cal X}\left( M\right) \Longleftrightarrow \nabla _{X}\overline{g}%
_{\epsilon }=0$ for all $\epsilon \neq 0$ and $X\in {\cal X}\left( M\right)
$)

\noindent Proof: It is obvious that $\nabla _{X}g=0$ and $\nabla _{X}\tau =0$
imply $\nabla _{X}\overline{g}_{\epsilon }=0.$ Now assume $\nabla _{X}%
\overline{g}_{\epsilon }=0$ for all $\epsilon \neq 0$. Indeed, it will
suffice to assume $\nabla _{X}\overline{g}_{\epsilon _{1}}=0$ and $\nabla
_{X}\overline{g}_{\epsilon _{2}}=0$ for two specific and distinct $\epsilon
_{1}\neq \epsilon _{2}$ (e.g. $\pm 1$). From $\nabla _{X}g+\epsilon
_{1}\nabla _{X}\left( \tau \otimes \tau \right) =0$ and $\nabla
_{X}g+\epsilon _{2}\nabla _{X}\left( \tau \otimes \tau \right) =0$ we get $%
\nabla _{X}g=0$ and $\nabla _{X}\left( \tau \otimes \tau \right) =0.$ By the
product rule
\begin{eqnarray*}
\left( \nabla _{X}\left( \tau \otimes \tau \right) \right) \left( Y,Z\right)
&=&\nabla _{X}\left( \left( \tau \otimes \tau \right) \left( Y,Z\right)
\right) -\left( \tau \otimes \tau \right) \left( \nabla _{X}Y,Z\right)
-\left( \tau \otimes \tau \right) \left( Y,\nabla _{X}Z\right) =0 \\
&=&X\left( \tau \left( Y\right) \tau \left( Z\right) \right) -\tau \left(
\nabla _{X}Y\right) \tau \left( Z\right) -\tau \left( Y\right) \tau \left(
\nabla _{X}Z\right) =0
\end{eqnarray*}
for all $X,Y,Z\in {\cal X}\left( M\right) .$ Taking $Y=Z=V$ we find $\tau
\left( \nabla _{X}V\right) =0$ by $\tau \left( V\right) =1.$ If we now take
in the above identity only $Z=V$ and use $\tau \left( \nabla _{X}V\right) =0$
we get
\[
X\left( \tau \left( Y\right) \right) -\tau \left( \nabla _{X}Y\right) =0
\]
which means $\left( \nabla _{X}\tau \right) \left( Y\right) =0,$ i.e. $%
\nabla _{X}\tau =0.$

\section{Conclusion}

\noindent We discuss here the relation between n-dimensional
pseudo-Riemannian manifolds with parallel vector fields and
Newton-Cartan structures. The latter  consist of two parallel
tensors, namely, a type-$(0,2)$ symmetric tensor $g$ that is
degenerate of corank-(n-1) and a non-vanishing 1-form $\tau$ such
that $g$ is non-degenerate on $ker(\tau)$ and a compatible
connection $\nabla$. From this data, it is possible to construct a
1-parameter family of pseudo-Riemannian metrics
${\bar{g}}_{\epsilon} = g + \epsilon \tau \otimes \tau$ whose
Levi-Civita connection coincide for any $\epsilon$ with the
(torsion-free) Newton-Cartan connection and the 1-form $\tau$ and
its dual vector $V$ are parallel. Conversely, from a
pseudo-Riemannian metric and a parallel vector field $V$ one may
reconstruct the Newton-Cartan structure making use of the dual
1-form $\tau$. We give formulas that characterize a metric $g$ and
a parallel vector field $V$ that are compatible with a
Newton-Cartan connection with torsion. Most of these results were
previously announced in an unpublished thesis \cite{limoncu}.

\section{Acknowledgements}

\noindent TD thanks Professors J. Ehlers and R. W. Tucker
for discussions that motivated this work.

\end{document}